\documentclass[aps,prl,twocolumn]{revtex4}%
\usepackage{amsfonts}
\usepackage{amsmath}
\usepackage{amssymb}
\usepackage{graphicx}%
\begin{document}
\title{Comment on \textquotedblleft Spin Correlations in the Paramagnetic Phase and
Ring Exchange in La$_{2}$CuO$_{4}$\textquotedblright}
\author{L. Raymond, G. Albinet}
\affiliation{L2MP, 49 rue Joliot Curie BP 146, Universit\'{e} de Provence, 13384 Marseille,
Cedex 13, France}
\author{A.-M. S. Tremblay}
\affiliation{D\'epartement de Physique and RQMP, Universit\'e de Sherbrooke, Sherbrooke,
Qu\'ebec, J1K 2R1, Canada.}
\keywords{Hubbard model, ring exchange, neutron scattering}
\pacs{75.40.Gb, 75.10.Jm, 75.25.+z}

\begin{abstract}
It is shown that the experiments of A.M. Toader, J. P. Goff, M. Roger, N.
Shannon, J. R. Stewart, and M. Enderle, Phys. Rev. Lett. \textbf{94}, 197202
(2005) do not provide definitive experimental evidence for ring exchange terms
in the Hamiltonian of La$_{2}$CuO$_{4}$, even though such terms may be present.

\end{abstract}
\maketitle

In a recent paper,\ Toader \textit{et al. }\cite{Toader} claimed to provide
definitive experimental evidence for ring exchange terms in the Hamiltonian of
La$_{2}$CuO$_{4}$ by comparing the experimental antiferromagnetic static spin
structure factor $S\left(  \mathbf{Q}\right)  $ with high-temperature series
expansion. Ring-exchange terms arise at intermediate coupling in the effective
low-energy theory for the Hubbard model. The ring-exchange parameters deduced
from Ref.\ \cite{Toader} agree with those found earlier\ \cite{Coldea}, and in
terms of the Hubbard model they correspond to $\kappa\equiv t/U=1/7.5$ with
$t=0.29%
%TCIMACRO{\unit{eV}}%
%BeginExpansion
\operatorname{eV}%
%EndExpansion
$.

\begin{figure}[ptb]
\begin{center}
\includegraphics[width=7.5cm]{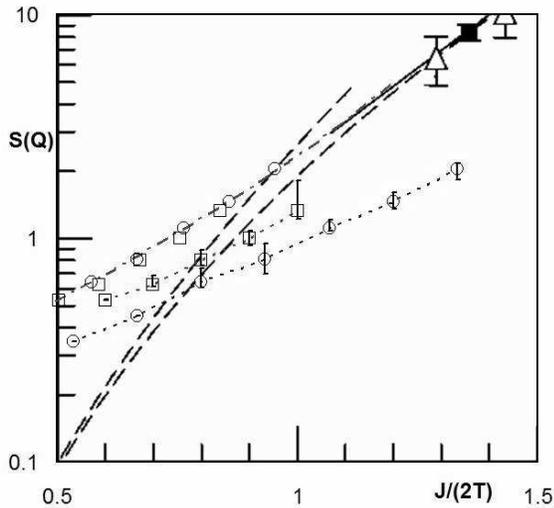}
\end{center}
\caption{High temperature series expansion from Ref.~\cite{Toader}, lower
dashed line is fifth order, higher dashed line is fourth order. Experimental
results in upper right-hand corner. Heisenberg model ~\cite{Elstner}
dashed-dotted line. QMC results $U=7.5t$ circles and $U=10t$ squares. When no
error bars are indicated, there is additional scaling, as discussed in the
text.}%
\end{figure}

Fig. 1 compares several results with the corresponding data in Fig.3 of
Ref.\ \cite{Toader}. The high-temperaure series expansion of $S\left(
\mathbf{Q}\right)  $ in Ref.\ \cite{Toader} are in units where
$J/t=2\widetilde{J}_{2}^{\left(  1\right)  }/t=4\kappa-64\kappa^{3}$ and they
are represented by the two dashed lines, along with three of the experimental
points in the upper right-hand corner. Our Quantum Monte Carlo (QMC) data for
$U=7.5t$ and $U=10t$ are given respectively by circles and by squares linked
by a dotted line as a guide to the eye. We used the determinental method with
discretization step $\Delta\tau=1/10.$ When the QMC data appear to have
reached a size independent value, we give as lower and upper error bars
statistical fluctuations on, respectively, the smallest and largest result.
The largest system size is typically $L\times L=12\times12$. When the size
dependence is still important, the upper error bar is an upper bound obtained
by a $1/L$ extrapolation of the values for the two largest system sizes.
Except for the lowest temperatures, the results have converged as a function
of $L$. The horizontal scale for QMC is obtained from $J/t=4t/U\equiv4\kappa$
with temperature in units of $t$, so that the scale $J/\left(  2T\right)  $
depends only on the ratio $\kappa$. We also show as a dash-dot line the value
of $S\left(  \mathbf{Q}\right)  $ obtained from high-temperature series
expansion of the Heisenberg model,\ \cite{Elstner} without ring exchange. We
take $\hbar=1$ for both the Heisenberg and QMC results so that the
corresponding $S\left(  \mathbf{Q}\right)  $ can be compared with QMC on an
absolute scale. However, the absolute scale of the results of
Ref.\ \cite{Toader} is not given, but on a logarithmic scale this is just a
shift of the origin that does not change the slope. So $S\left(
\mathbf{Q}\right)  $ for the Heisenberg model and for QMC have all been
shifted by precisely the same value to compare with Ref.\ \cite{Toader}.
Clearly, as expected, the larger the value of $U$ in the QMC calculation$,$
the better the agreement with the Heisenberg model. Note however that both
values of $U$ are much closer to the Heisenberg result than to the results of
Ref.\ \cite{Toader}, despite the fact that $U=7.5t$ should correspond to the
exchange parameters (including ring) used in that reference.

In Ref.\ \cite{Toader} it was argued that agreement with experiment was better
with ring exchange than without it because the high-temperature series results
dove tail better with the experiment than the results obtained without ring
exchange. We have here a counter example since with just a shift of the origin
of the vertical axis of either the $U=7.5t$ or $U=10t$ results of the Hubbard
model we can smoothly join the experimental data. As an added observation, we
show in Fig.\ 1 that when our QMC data (represented by the same symbols as
before but this time without error bars) are plotted as a function of
$J/t=2\widetilde{J}_{2}^{\left(  1\right)  }/t=4\kappa-64\kappa^{3}$, as
suggested in Ref. \cite{Toader}, then they fall close to the Heisenberg curve
and they all join smoothly the experiment.

The experimental results on $S\left(  \mathbf{Q}\right)  $ are in the
universal regime that can be described by the non-linear sigma model and hence
they are insensitive to microscopic details, as noted in Ref.\ \cite{Toader}.
Our results suggest that unless this $S\left(  \mathbf{Q}\right)  $ can be
measured in the non-universal regime at higher temperature, it cannot lead to
an accurate value of $U/t$ (and hence of ring exchange contribution) by a
smoothness argument, especially if additional parameters such as
next-nearest-neighbor hopping are taken into account. Experiments involving
measurement of $S\left(  \mathbf{Q,}\omega\right)  $ at higher energies and
shorter wavelengths are more informative, as discussed in Ref.\ \cite{Coldea},
but even in that case it is important to also take into account detailed
information on the band structure, including further neighbor hoppings, to
obtain reliable values of $U$ and consequent ring exchange terms$.$%
\cite{Delannoy}

We are indebted to M. Gingras, J.-Y. Delannoy, M. Roger and N. Shannon for
useful conversations.

\end{document}